\documentclass[aps,prl,twocolumn,superscriptaddress,amsmath,amssymb,floatfix,showpacs]{revtex4}
\usepackage[dvips]{graphics}
\usepackage{graphicx}
\usepackage{amsfonts}
\usepackage{amssymb}
\usepackage{amsmath}
\usepackage{subfigure}
\usepackage{wasysym}
\usepackage{color}
\DeclareMathAlphabet{\mathpzc}{OT1}{pzc}{m}{it}
\begin{document}

\title{Stable Vortex-Bright Soliton
Structures in Two-Component Bose Einstein Condensates}
\author{K.\ J.\ H.\ Law}
\affiliation{Warwick Mathematics Institute, University of Warwick, Coventry CV4 7AL, UK}
%\affiliation{Department of Mathematics and Statistics,
%University of Massachusetts,
%Amherst MA 01003-4515, USA}
\author{P.\ G.\ Kevrekidis}
\affiliation{Department of Mathematics and Statistics,
University of Massachusetts,
Amherst MA 01003-4515, USA}
\author{Laurette S. Tuckerman}
\affiliation{PMMH-ESPCI, CNRS (UMR 7636), Univ. Paris 6 \& 7,
 75231 Paris Cedex 5, France}

\begin{abstract}

We report the numerical realization of robust 2-component structures in 2d and
3d Bose-Einstein Condensates with non-trivial topological charge in one 
component.  We identify a stable symbiotic state in which a
higher-dimensional bright soliton exists even in a homogeneous setting with
defocusing interactions, due to the effective potential created by a stable
vortex in the other component. The resulting vortex-bright solitary waves,
generalizations of the recently experimentally observed dark-bright solitons,
are found to be very robust in both in the homogeneous medium and in the
presence of parabolic and periodic external confinement.
\end{abstract}

\maketitle

{\it Introduction}. Vortices in nonlinear field theory have a time-honored
history \cite{Pismen}. They are among the most striking features of
superfluids, play a role in critical current densities and resistances of
type-II superconductors through their transport properties, and are associated
with quantum turbulence in superfluid helium~\cite{donnelly}.  The advent of
Bose-Einstein condensates (BECs) 15 years ago \cite{cornell,ketterle1} has
produced an ideal setting for exploring relevant phenomena.  Since the
experimental observation of matter-wave vortices ~\cite{Matthews99}, by using
a phase-imprinting method between two hyperfine spin states of a $^{87}$Rb BEC
\cite{Williams99}, the road opened for an extensive examination of vortex
formation, dynamics and interactions.  Stirring the BECs \cite{Madison00}
above a certain critical angular speed \cite{Recati01,Sinha01,corro,Madison01}
led to the production of few vortices \cite{Madison01} and even of very robust
vortex lattices \cite{Raman}. These structures have been produced by other
experimental techniques, such as dragging obstacles through the BEC
\cite{kett99} or the nonlinear interference of condensate fragments
\cite{BPAPRL}. Later, not only unit-charged, but also higher-charged
structures were produced \cite{S2Ket} and their dynamical (in)stability was
examined. This field also has strong similarities and overlap with the
emergence of vortices and even vortex lattices in nonlinear optical settings;
see e.g.~\cite{yuripismen,dragomir}.

\begin{figure}[tbh]
\begin{center}
\includegraphics[width=0.48\textwidth]{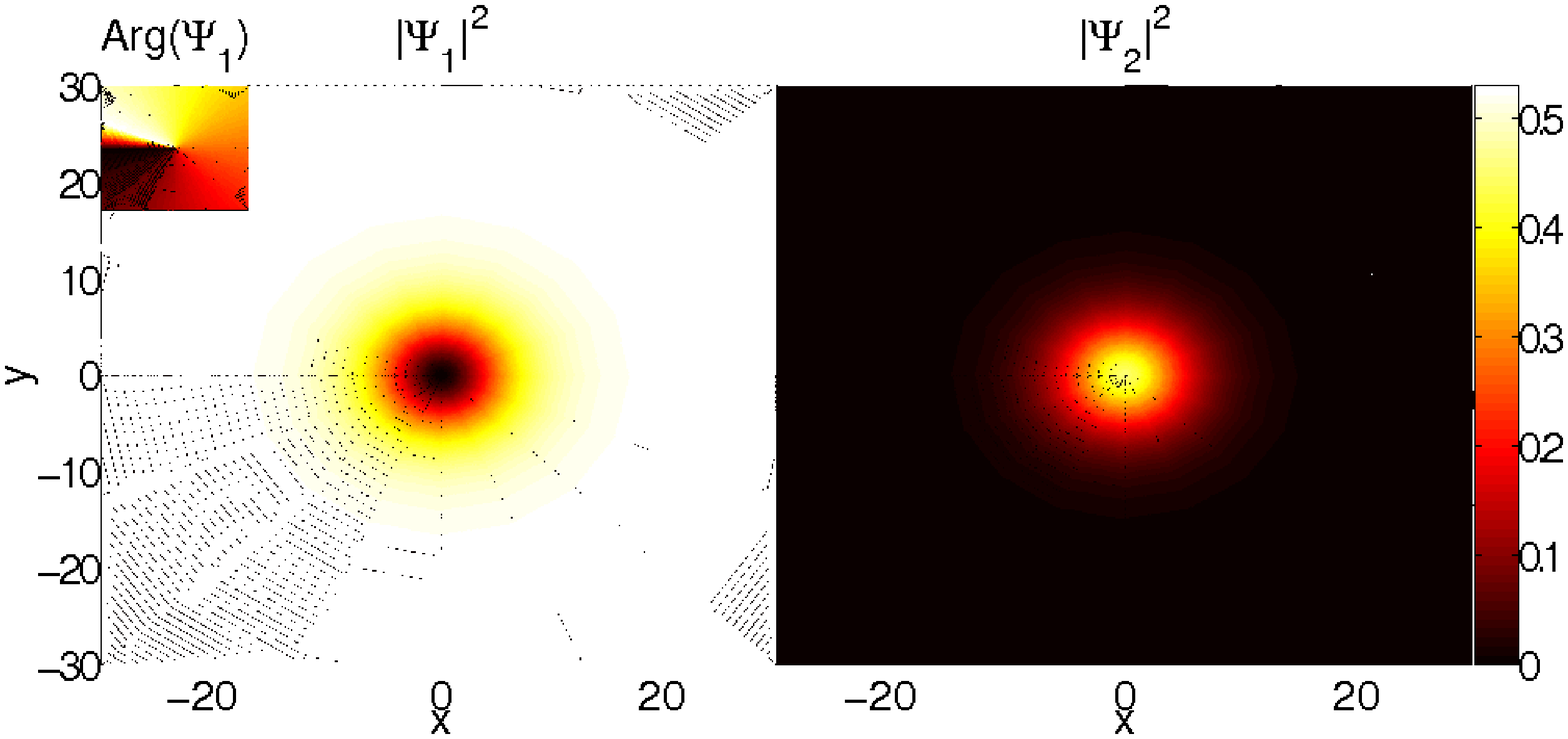}
\includegraphics[width=0.22\textwidth]{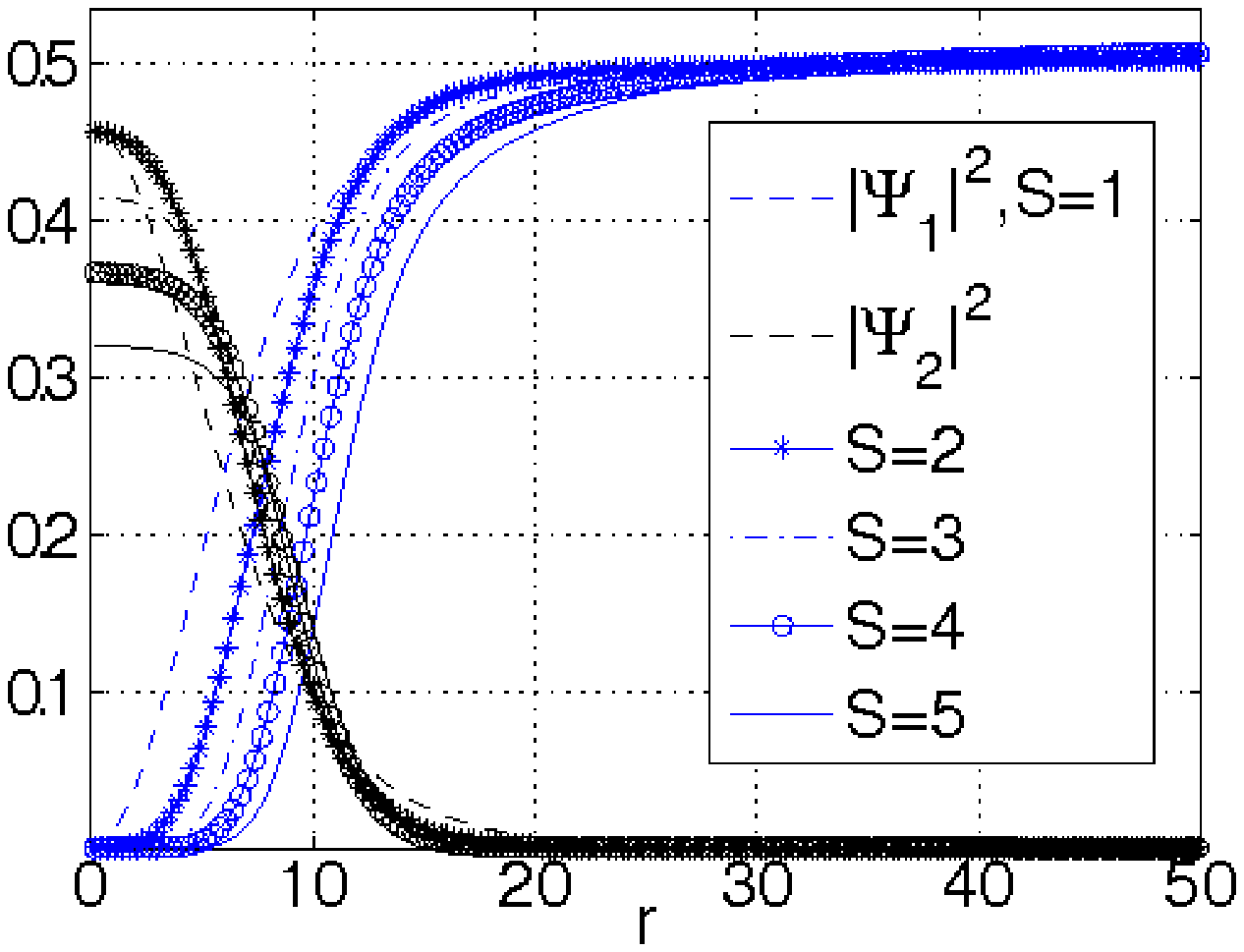}
\includegraphics[width=0.22\textwidth]{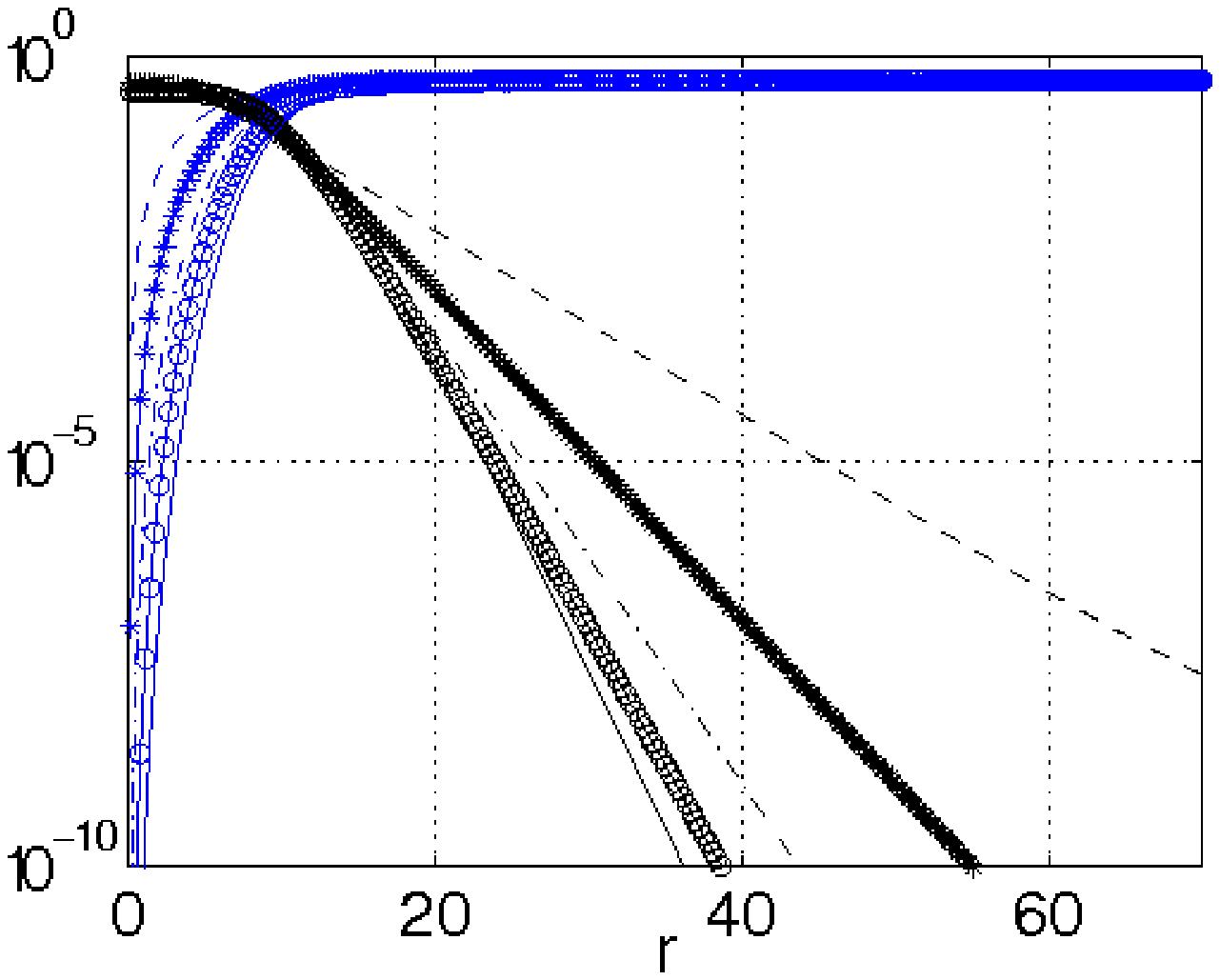}
\end{center}
\caption{(Color online) The energetically stable S=1 vortex-bright without
  external potential for $R=0.99$, $r_{\rm max}=60$ and $N=5900$ (top row).
  The bottom row shows radial profiles of unit and higher charge vortex-bright
  solitons in a homogeneous medium on a regular (left) and logarithmic (right)
  scale.  All profiles are for $N=10000$, $R=0.99$, and $r_{\rm max}=80$.}
\label{fig1}
\end{figure}

Another remarkable possibility in both
BEC~\cite{Myatt1997a,Hall1998a,Stamper-Kurn1998b} and in nonlinear optics,
e.g.~\cite{stiegl}, is that of multi-component settings.  Matter waves exhibit
rich phase separation dynamics driven by the nonlinear interatomic
interactions between different species or states that make up the BECs.
Longitudinal spin waves~\cite{Lewandowski2002a}, transitions between
triangular and interlaced square vortex lattices~\cite{Schweikhard2004a},
striated magnetic domains~\cite{Miesner1999a,Stenger1999a}, and robust target
patterns~\cite{hall07} have all been observed, as well as tunable interspecies
interactions \cite{minardi} and transitions between miscible and immiscible
dynamics \cite{papp}.

We interweave these two settings, motivated by \cite{hamburg} in which
dark-bright solitons have been created in a 
%LST 1D --> 1d for consistency
%quasi-1D, 
quasi-1d
two-component
BECs. These structures were predicted \cite{buschanglin} and extended to more
complex settings such as spinor condensates \cite{DDB} (as dark-dark-bright,
or dark-bright-bright solutions), but were only realized experimentally in
2008. These are often termed ``symbiotic solitons'', as the bright component
would be impossible to sustain under repulsive inter-atomic interactions
(i.e., defocusing nonlinearities, as considered here), unless the
dark-component creates an ``effective potential'', of which the bright soliton
is a bound state.  
The coupled bright solitary waves  \cite{vpg} and the
gap ones  of \cite{adhikari} constitute additional examples of symbiotic 
%LST
%structures in their respective settings.
structures.
We consider higher-dimensional realizations
\cite{buschanglin} 
%LST
%in the form of vortex-bright 
i.e.~vortex-bright
solitons of various
topological charge in 2d, as well as in 
%LST 3 --> 3d
%3 
3d~\footnote{
%LST
%We restrict our considerations to repulsive interactions where such 
%structures are symbiotic; in attractive interaction settings, 
%where such states can be self-trapped and with a vanishing tail, 
We consider only repulsive interactions, where such structures are
  symbiotic. With an attractive interaction, where such states
can be self-trapped with a vanishing tail, they were originally
proposed in Z.H. Musslimani {\it et al.}, 
Phys. Rev. Lett. {\bf 84}, 1164 (2000).}.  We find these symbiotic
configurations to be robust, with or without parabolic external confinement.
In an optical lattice, the unstable vortex may in fact be stabilized by the
bright soliton. The stability persists in 3d, while for traps elongated in the
direction of the vortex core, additional negative energy (potentially
instability bearing) modes \cite{pu} emerge, as in the single-component vortex
\cite{feder}. The work of~\cite{Matthews99} has already offered a
prototypical dynamical realization of such states (analogous to their quasi-1d
counterparts of ~\cite{buschanglin} by ~\cite{hamburg}) and attests to
their experimental relevance.  We will first give the physical setup, then
discuss the numerical methods and lastly display the results, as well as
future directions.

%\begin{figure}[tbh]
%\begin{center}
%\includegraphics[width=0.22\textwidth]{higher_radialBcrop.eps}
%\includegraphics[width=0.22\textwidth]{higher_radial_logBcrop.eps}
%\end{center}
%\caption{(Color online) Radial profiles of unit and higher charge vortex-bright
%solitons in a homogeneous medium on a regular (left)
%and logarithmic (right) scale.  All profiles are for $N=10000$,
%$R=0.99$, and $r_{\rm max}=80$.}
%\label{fig2}
%\end{figure}

{\it Physical Setup}. The 
non-dimensional Hamiltonian for a two-component condensate in the mean-field
approximation reads \cite{pethickpit}:
\begin{equation}
H = \int d{\bf r} (\nabla \Psi)^{\dagger}(\nabla \Psi) +
\Psi^{\dagger} V({\mathbf r}) \Psi +
\frac{1}{2} |\Psi|^{2\dagger} U |\Psi|^2 - \Psi^{\dagger} M \Psi
\label{hamiltonian}
\end{equation}
\noindent where $\Psi({\mathbf r}) \in \mathbb{C}^2$ is the pseudo-spinor
order parameter,
$|\Psi|^2=(|\Psi_1|^2,|\Psi_2|^2)^{\dagger}$,
$M={\rm diag} \{\mu_1, \mu_2\}$ is the diagonal matrix of chemical potentials
associated with the conservation of the number of atoms
%Lagrange multipliers associated to the following constraints
$N_1 = \int d{\bf r} |\Psi_1|^2$ and
%& \hspace{1 cm}
$N_2 = \int d{\bf r} |\Psi_2|^2$; a related useful diagnostic
is $R = N_1 / (N_1+N_2) = N_1 / N$.
$U$
%with det$(U)=|U|<0$
is a $2\times2$ matrix accounting for the
effectively nonlinear interatomic interactions.
For the
$|1,-1\rangle$ and
$|2,1\rangle$ components of $^{87}$Rb
%, henceforth denoted
%$|1\rangle$ and $|2\rangle$,
we can use \cite{hall07} $U_{11}=1.03$, $U_{12}=U_{21}=1$ and $U_{22}=0.97$.
These determine, through the negative sign of det$(U)=|U|$, the
immiscible nature of the interactions leading to phase 
separation~\cite{Hall1998a,hall07}.  
The confining potential is
\begin{equation}
V({\mathbf r}, z) = \underbrace{\frac{\omega_r^2}{4}
|{\mathbf r}|^2 + \frac{\omega_z^2}{4}z^2}_{V_{\rm MT}} +
\underbrace{A[\sin^2(2 \omega_r x)+\sin^2(2 \omega_r y)]}_{V_{\rm OL}},
\label{potential}
\end{equation}
\noindent where $V_{\rm MT}$ is the parabolic component (often created
magnetically) and $V_{\rm OL}$ the periodic (optical) lattice component.  
The time and length scales are $1/\omega_n$ and $\sqrt{\hbar/m\omega_n}$, 
where $m$ is the atomic mass 
and $\omega_n$ is an arbitrary frequency in Hz.  For $^{87}$Rb with scattering
length of $a_{12}=5.5$nm, $(\omega_r,\omega_z)=2 \pi \times (8,40)$ Hz, and
choosing $\omega_n=5/4\omega_z$, the ratio between the actual and
non-dimensional number of atoms is $N_{{\rm fac},3d} =
(\hbar/2m\omega_n)^{3/2}(\hbar\omega_n/g_{3d})=10$, where $g_{3d}=(4\pi
\hbar^2 a_{12}/m)$ is the dimensional interaction parameter. For a 
2d reduction, the interaction parameter is $g_{2d}=g_{3d}(m\omega_z/2\pi
\hbar)^{1/2}$ (e.g.~\cite{jaksch}) and taking
$\omega_n=\omega_r$, the amplification factor is $N_{{\rm fac},2d}=30$.  The
equations of motion $(\dot\Psi,c.c.)^T=J \sigma (\delta
H/\delta\Psi,c.c.)^T=J \sigma D H$,
%$\dot\Psi=J DH$
where $J={\rm diag}(-iI,iI)$ and $\sigma$ interchanges rows $(3,4)$ with
$(1,2)$, for this infinite-dimensional Hamiltonian system are
\begin{equation}
i \dot{\Psi} = - \nabla^2 \Psi + V({\mathbf r}) \Psi + U |\Psi|^2 \cdot \Psi
- M \Psi,
\label{gpe}
\end{equation}

The stability of stationary solutions is determined by the eigenvalues of
the Hessian of the Hamiltonian, $\sigma D^2 H$, and of $J \sigma D^2 H$.
Negative eigenvalues of
$\sigma D^2 H$ indicate {\it energetic instability}, since ``dissipative''
perturbations (e.g.~from exchanges of atoms with the thermal cloud if the
temperature deviates from zero) in the system can render them {\it dynamically
  unstable}, as can collisions with other eigendirections even in the pure
Hamiltonian (zero-temperature) system.  The linear stability of the latter
system is examined through the eigenvalues $\lambda=\lambda_r + i \lambda_i$
of $J \sigma D^2 H$; instability arises when $\lambda_r\neq 0$ since, due to
the Hamiltonian structure, the eigenvalues are symmetric over both the real
and imaginary axes.  From prior experience which is confirmed
again here, linear stability indicates evolutionary {\it non-linear stability}
in the mean-field model, at least for time scales on the order of tens of
seconds (this is not generically the case for non-linear static solutions of
Hamiltonian systems).

The variation can be posed in the $\{\Psi,\Psi^*\}$ or the $\{\Psi_{\rm
  real},\Psi_{\rm imag}\}$ basis.  The former is useful when the
potential is axisymmetric, since then small excitations to a stationary
solution $\Psi=(\Psi_1(r) e^{i S_1 \theta},\Psi_2(r) e^{i S_2 \theta})$ of the
form $\psi = (a_1,a_2)^T({\mathbf r}) e^{\lambda t} + (b_1,b_2)^T({\mathbf r})
e^{\lambda^* t}$ will have definite angular momentum
$\alpha_j(r,\theta) = \tilde{\alpha}_j(r) e^{i \kappa_{\alpha_j}\theta}$.  If
we set $\kappa_{a_1}=\kappa$ then $\kappa_{b_1}=\kappa-2S_1$,
$\kappa_{a_2}=\kappa-S_1+S_2$, and $\kappa_{b_2}=\kappa-S_1-S_2$, so a single
index $\kappa$ will indicate the angular momentum of the excitation with given
eigenvalue $\lambda$.  Hence, the spectrum of eigenvalues $\{\lambda\}$ can be
decomposed as the union of the spectra $\{\lambda_{\kappa}\}$ pertaining 
to angular momentum $\kappa$.
%The excitations have the symmetry
%$\{\lambda,a,b\} \rightarrow \{\lambda^*,b^*,a^*\}$.
We will also assume $S_2=0$, so that $S_1=S$ and
$\kappa_{a_2}=\kappa_{b_2}$.  It has been shown numerically \cite{pu} and
analytically \cite{kollar} that instability windows arise in
a single component with topological charge $S$ only for wave numbers with
$|\kappa| < S$.
%$-S < \kappa < 3 S$.  Due to the
%relations and symmetry mentioned above this is equivalent to
%$-S < \kappa S$.
The null eigenvalues corresponding to gauge invariance appear in the spectrum
of $\kappa=S$.  For a single component in a parabolic trap, an
anomalous mode for $\kappa=S-1$ 
converges to zero as $\omega_r \rightarrow 0$, accounting for
translational invariance and leading to the energetic stability of the $S=1$
vortex without external potential.  For each $0 \leq \kappa < S-1$ 
($S>1$) an anomalous mode leads to windows of instability
\cite{pu,kollar}.  We show that these can be significantly
suppressed, although the $S-1$ spectrum occasionally leads to small
instability windows for a small fraction bright-soliton component, $N_2 \ll
N_1$, for large $N$ with parabolic trap (no windows were observed without
the trap).

%The windows of instability characteristic of the $S-2$
%spectrum for $R=1$ disappear for smaller $R$ (the right panel
%depicts this for the case $S=2$).}

{\it Numerical Methods}. Our methods extend those in, e.g.,~\cite{huepe,todd}.
The spatial discretization in $(r,\theta,z)$ employs Chebyshev polynomials to
represent $r$ dependence ~\cite{trefethen}.  The Fourier modes representing 
$\theta$ and $z$ make the Laplacian operators diagonal in these
directions.  To identify stationary states of (\ref{gpe}), we first obtain an
initial estimate via imaginary-time (i.e., replacing $t \rightarrow it$)
integration using a first-order implicit/explicit Euler scheme with $\Delta t
= 10^{-2}$.  We then refine the solution 
using Newton's method.  The linear system arising at each Newton 
step is solved using the matrix-free IDR(s) algorithm
\cite{sonneveld,gijzen},
which requires only the action of the Hessian. To accelerate
inversion, we precondition the system with the inverse Laplacian,
using its block diagonal structure.  Hence, we solve the system
$\nabla^{-2} D^2 H(\Psi_n) \Delta_n = \nabla^{-2} D H(\Psi_n)$ and update
$\Psi_{n+1}=\Psi_n-\Delta_n$ for $n=0,1,\ldots$.
Fewer than 5 Newton iterations usually achieve an accuracy of
$||\nabla^{-2}D H(\Psi)||_{l^2}/||\Psi||_{l^2}<10^{-12}$.

\begin{figure}[tbh]
\begin{center}
\includegraphics[width=0.21\textwidth]{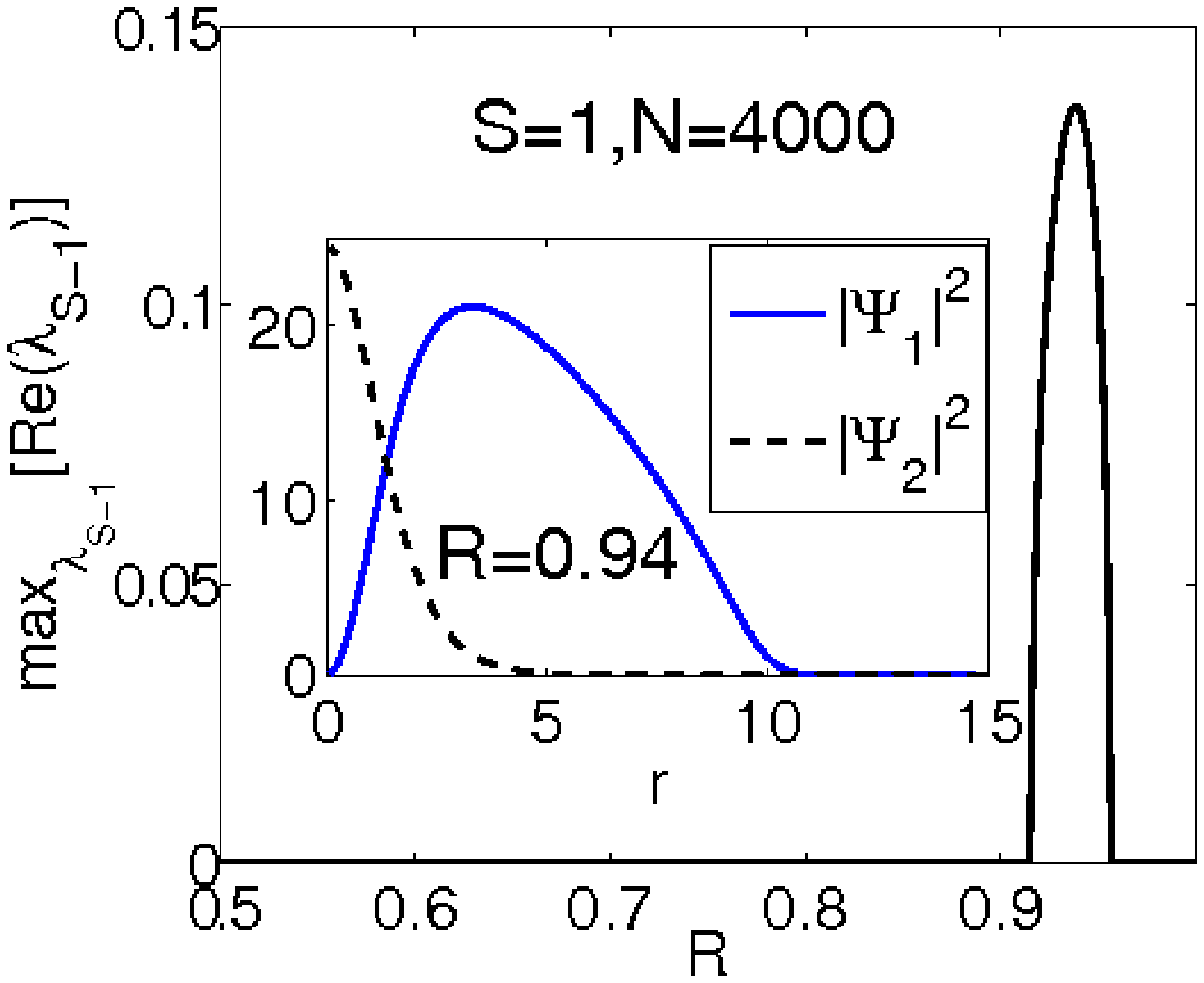}
\includegraphics[width=0.24\textwidth]{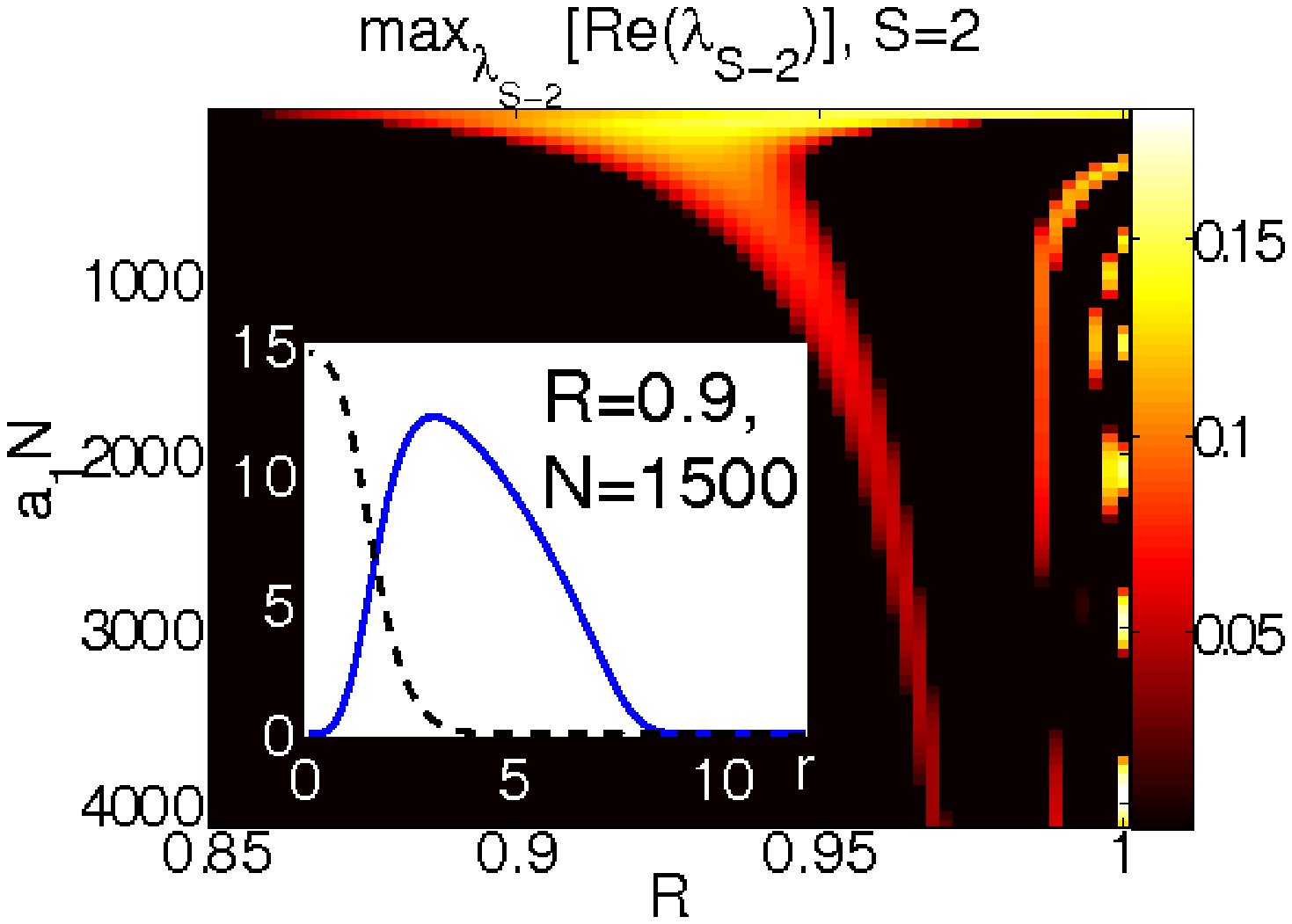}
\end{center}
\caption{(Color online) Growth rate of the $S-1$ mode
as a function of $R$ with $N=4000$ for $S=1$ (left),
and growth rate of the $S-2$ mode as a function of
$R \times a_1N$ for $S=2$ (right).}
\label{fig3}
\end{figure}

For each stationary solution $\Psi$, we use the matrix-free Implicitly
Restarted Arnoldi algorithm to iteratively compute the eigenpairs of
the linearization 
$J \sigma
D^2 H(\Psi)$ to a specified tolerance \cite{sorenson}.  
In order
to find the desired eigenvalues we use inverse iteration, with 
the IDR(s) method and
inverse Laplacian preconditioning to solve the linear systems, as above.
Here, the preconditioner is taken to be $[J \sigma (\nabla^2)]^{-1}$, so that
each iteration solves $\nabla^{-2}D^2 H(\Psi) v_{n+1} = -\nabla^{-2} \sigma J
v_n$.
%\footnote{To compute the eigenvalues of the Hessian, we precondition
%with $\nabla^{-2}$ as in Newton's method.}

We used a resolution in $(r,\theta,z)$ of $40 \times 64 \times
80$ to represent non-axisymmetric solutions and eigenvectors.  For
axisymmetric solutions, quantitative accuracy requires only 30 radial modes
for $N<1000$, but up to 200 modes for larger $N$.  For eigenvectors, we use
only $S+1$ modes in $\theta$ (see introduction) and identify quantitatively
all expected invariant and negative directions and windows of instability from
\cite{pu}.

{\it Results}. 
In 2d ($\omega_z \rightarrow \infty$) for $\omega_r=A=0$, we
first demonstrate in Fig. \ref{fig1} the existence of an {\it energetically
  stable} (and hence also dynamically stable) vortex-bright soliton state.
This is so for all of the $R-N$ values that we have sampled.  Notice the
symbiotic nature of the state, as a bright soliton would be impossible to
support under the repulsive/self-defocusing interactions considered herein.
Indeed, a similar state exists for vortices of higher topological charge $S$,
as shown in Fig. \ref{fig1} for $S \leq 5$.  The logarithmic scale
shows that the soliton is more localized for larger $S$.  In this
case, the negative energy modes \cite{pu} in the spectra associated
to $0 \leq \kappa<S-1$ may lead to dynamical instability from complex quartets
of eigenvalues as a result of Hamiltonian-Hopf bifurcations which can occur
upon collision with positive modes \cite{meer}.

\begin{figure}[tbh]
\begin{center}
\includegraphics[width=0.48\textwidth]{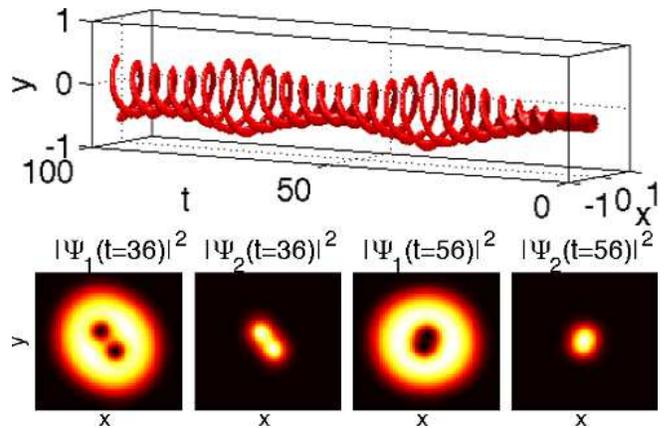}
\end{center}
\caption{(Color online) The evolution in time of the unstable $S=2$
solution for $(R,N)=(0.93,200)$ perturbed in the direction of the growing excitation.
Top: approximate vorticity density iso-contours show trajectories 
of the two vortices. Bottom: first splitting and rejoining.}
\label{fig3.5}
\end{figure}

In a parabolic trap, the $S=1$ vortex-bright structure
remains dynamically stable.  However, breaking of
translational invariance produces a negative energy mode in the
$\kappa=S-1$ spectrum.  
Thus, 
an additional control parameter (through the atom number
of the bright component) may lead to collisions of this mode with 
positive energy modes and, hence, rare isolated windows of instability 
arising for large $R<1$ and large $N$; 
%LST
%see the left panel of Fig. \ref{fig3} for an example of 
see Fig. \ref{fig3} (left) for 
such a window for $S=1$ as a function of $R$ when $N=4000$.
A
similar feature has recently been shown in the
dark-bright
1d
analog of the vortex-bright states~\cite{middelk}.

For higher-charge vortex-bright structures, the same situation holds for the
$\kappa=S-1$ spectrum, while for $R=1$ windows of
instability arise from the negative modes in the spectra of $0 \leq
\kappa<S-1$.  As $R$ decreases, however, these windows of instability are
generically suppressed by the increasing presence of the second bright
component. Fig. \ref{fig3} (right) depicts the growth rate of the $S-2$
spectrum for $S=2$ over $R-N$ parameter space.  An example of an unstable
solution perturbed in the growing excitation direction is depicted in
Fig. \ref{fig3.5}.  The vortices first split from the center, and begin
to part and precess, but once they are far enough and
the bright component is bimodal, they approach again and the
bright component resumes uni-modality.
The sequence repeats, similarly to single-component $S=2$ vortices
\cite{nilsen}.

When we impose an additional sinusoidal lattice potential, $A > 0$,
the one-component ($R=1$) $S=1$ vortex may become unstable
(due to resonant eigenvalue collisions and ensuing oscillatory 
instabilities), at least for $A$
sufficiently large \cite{ours_yannis}.  The same holds
for a large mass ratio $R<1$.  However, below a critical $R$,  
once again the bright component has a stabilizing influence.

\begin{figure}[tbh]
\begin{center}
\includegraphics[width=0.48\textwidth]{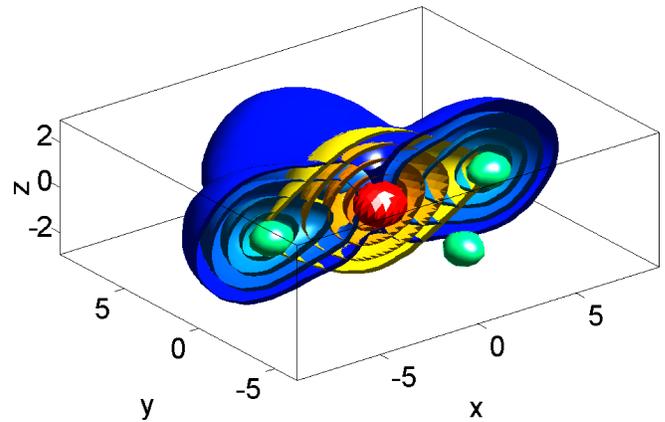}
\end{center}
\vspace*{-0.5cm}
\caption{(Color online) Iso-contours of the density of the stable symbiotic
vortex-bright structure in the presence of an optical lattice with
$\omega_z=5\omega_r$, $R=0.5$ and $N=40$.
Vortex surfaces are blue-scale (darker) and soliton
surfaces are yellow-scale (lighter).  This 3d stationary
state is stabilized by the second component,
%as one can see from the
which displaces the vortex component at its core.}
%by the other component.}
\label{fig4}
\end{figure}

The vortex-bright structure is stable 
in 3d without the trap and with periodic boundary conditions in $z$.
Indeed this is immediately clear 
upon Fourier transforming in $z$, since the spectrum of the Hessian decouples 
into an infinite family of sub-spectra equal to the 2d spectra shifted 
by $k_z^2$, and hence it remains non-negative.
It is stable in the trapped case as well for $A=0$ and $\omega_z =
5\omega_r$, and indeed for $\omega_z>\omega_r$.  When $\omega_z= \omega_r$,
the solution has 
another rotational invariance, and additional negative
energy modes emerge for $\omega_z < \omega_r$.  
For $2 \omega_z = \omega_r$
%LST But this is in opposite direction, i.e. $\omega_r > \omega_z$
there are at least two additional negative energy modes, although this 
may not
lead to dynamical instability.  
%LST What does ``addition'' of the 3d lattice mean?
%LST Would it be possible to replace this phrase by ``In 3d''?
Upon addition of 
%LST
the lattice, $A>0$,%in 3d 
%the 3d lattice
%In 3d
the results are expected to be similar to 2d 
(up to 
%LST
%the 
considerations
of the aspect ratio of the harmonic trapping).  See
Fig. \ref{fig4} for an example with $\omega_z = 5\omega_r$.

{\it Discussion}. We have generalized the dark-bright, quasi-1d soliton that
has been predicted theoretically and observed experimentally in
BECs to that of a vortex-bright robust dynamical entity that emerges as a
stable structure in both 2d and 3d condensates (although similar concepts
could be directly applicable to the nonlinear optics of defocusing optical
media). We also examined relevant structures in the presence of parabolic
(magnetic) and periodic (optical) trapping and found that they remain
stable. While instabilities may arise (e.g.~for higher
topological charge, $S$, or as a result of the lattice), these are usually
alleviated/suppressed by the presence of the second component in 2d and
3d.

It would be interesting to determine the robust existence of such
waveforms, which are well within the reach of recent experiments,
e.g.~\cite{papp,hall07}.  Our study suggests that higher-charge vortices
in a single component may be stabilized by an external blue-detuned laser-beam
potential acting as the bright-soliton here. Hence, the stability of 
such vortices should be systematically examined in the
presence of external potentials. Other themes such as
multi-vortex-bright soliton interactions and lattices would also be natural
extensions of the present work.

%On the other hand, it would be
%appealing and relevant to examine to what extent symbiotic
%structures could be further generalized in the richer three-dimensional
%setting, by considering symbiotic states consisting of bright solitons with
%more complex entities such as vortex rings; this could constitute a stable
%dynamical example of a ring-bright soliton (coupled to a vortex-ring
%in the other component).

{\it Acknowledgments}. PGK gratefully acknowledges support from NSF-DMS-0349023,
0806762 and the Alexander von Humboldt Foundation.
%KJHL %gratefully
%acknowledges D.~Barkley and A.~Aftalion
%for useful discussions.

\end{document}